
\input phyzzx
\nopubblock
\sequentialequations
\twelvepoint
\overfullrule=0pt
\tolerance=5000

\REF\tolman{R. C. Tolman, {\it The principles of Statistical Mechanics},
(Dover, New York, 1979) }

\REF\bombelli{L. Bombelli, R. Koul, J. Lee, and R. Sorkin,
Phys. Rev.  {\bf D34} (1986) 373.}

\REF\thesis{C. Holzhey, {\it Princeton University thesis}, 1992,
(unpublished)}

\REF\srednickione{M. Srednicki,
Phys. Rev. Lett. {\bf 71}, (1993) 666 }

\REF\ccfw{C.G.Callan and F. Wilczek,
{\it On Geometric Entropy},
hep-th-9401072}

\REF\dowker{J.S.Dowker,
{\it Remarks on Geometric Entropy},
MUTP-94-2,hep-th-9401159}

\REF\kabat{D. Kabat and M.J.Strassler,
{\it A Comment on Entropy and Area},
RU-94-10,hep-th/9401125}

\REF\susskindfour{L. Susskind,
{\it Some Speculations About Black Hole Entropy in String Theory},
RU-93-44,hep-th/9309145}

\REF\srednickitwo{M. Srednicki, {\it private communication}}

\REF\preskill{J. Preskill, seminar at Princeton University}

\REF\keski{E. Keski-Vakkuri and S. Mathur,
{\it Evaporating Black Holes and Entropy},
MIT-CTP-2272,hep-th/9312194}

\REF\ginsparg{P. Ginsparg,
{\it Applied Conformal Field Theory},
in {\it Fields, Strings, and Critical Phenomena}
Les Houches, Session XLIX, 1988, eds. E.Br\'{e}zin and J. Zinn-Justin.}

\REF\cardy{J. L. Cardy,
{\it Conformal Invariance and Statistical Mechanics},
in {\it Fields, Strings, and Critical Phenomena}
Les Houches, Session XLIX, 1988, eds. E.Br\'{e}zin and J. Zinn-Justin.}

\REF\birrell{N. D. Birell and P. C. W. Davies, {\it Quantum
Fields in Curved Space} (Cambridge University Press, Cambridge, 1982).}

\REF\carlitz{R. D. Carlitz and R. S. Willey. Phys.Rev. {\bf D36}
(1987), 2327, Phys.Rev. {\bf D36} (1987), 2336.}

\REF\chung{T-D. Chung and Herman Verlinde,
{\it Dynamical Moving Mirrors and Black Holes},
PUPT-1430,hep-th/9311007}

\line{\hfill }
\line{\hfill PUPT 1454, IASSNS 93/88}
\line{\hfill hep-th/9403108}
\line{\hfill March 1994}

\titlepage
\title{Geometric and Renormalized Entropy in Conformal Field Theory}

\author{Christoph Holzhey}
\vskip.2cm
\author{Finn Larsen}
\centerline{{\it Department of Physics }}
\centerline{{\it Joseph Henry Laboratories }}
\centerline{{\it Princeton University }}
\centerline{{\it Princeton, N.J. 08544 }}
\vskip .2cm
\author{Frank Wilczek\foot{Research supported in part by DOE grant
DE-FG02-90ER40542.~~~wilczek@iassns.bitnet}}
\vskip.2cm
\centerline{{\it School of Natural Sciences}}
\centerline{{\it Institute for Advanced Study}}
\centerline{{\it Olden Lane}}
\centerline{{\it Princeton, N.J. 08540}}
\endpage

\abstract{In statistical physics, useful notions of
entropy are defined with respect to
some coarse graining procedure over a microscopic model.  Here we
consider some special problems that arise when the microscopic model
is taken to be relativistic quantum field theory.  These problems are
associated with the existence of an infinite number of degrees of freedom
per unit volume.  Because of these the microscopic entropy can, and
typically does, diverge for sharply localized states.
However the difference in the entropy between two
such states is better
behaved,
and for most purposes it is the useful quantity to consider.  In particular,
a renormalized entropy can be defined as the entropy relative to the ground
state.  We make these remarks quantitative and precise in a simple model
situation: the states of a conformal quantum field theory
excited by a moving mirror.  From this work, we attempt to draw some
lessons concerning the ``information problem'' in black hole physics.}

\endpage

\chapter{Introduction}

Despite many startling experimental discoveries and
revolutionary changes in the foundations of theoretical physics,
the principles of thermodynamics have remained essentially unchanged since
their formulation by Carnot and Clausius in the early nineteenth century.
The reason for this unique stability has, in broad terms, long been
known.  It is that the thermodynamic laws are statistical regularities
among coarse-grained, essentially macroscopic, quantities, which follow
under very general assumptions about the underlying microscopic dynamics.
A particularly clear, detailed exposition of this circle of ideas can be
found in Tolman's classic book [\tolman ].

Nevertheless the derivation of macroscopic thermodynamics from microscopic
dynamics is not {\it a priori}, and one must examine it critically in the
light of changes in our understanding of the microscopic dynamics.
In particular, in relativistic
quantum field theory there are in principle an infinite number of
degrees of freedom per unit volume, and questions arise whether all these
degrees of freedom can come into equilibrium in a finite time, and whether
one encounters ultraviolet divergences in thermodynamic quantities
(and if so whether they may be
regulated and renormalized).

This tension becomes acute when one discusses the application of
thermodynamics to space-time geometries containing
black hole event horizons  and to
the closely related moving mirror
model.   For in these situations there is, as we shall discuss in detail
below, effectively a {\it sharp\/} boundary between accessible and
inaccessible regions of space-time (for a natural class of observers).
In the presence of
such a sharp boundary, the aforementioned ultraviolet
divergences actually do arise.

Our goal in this paper is to discuss these issues concretely in the
simplest possible non-trivial setting, that of conformally invariant
field theory in 1+1 space-time dimensions.
In the following Section 2 we shall
develop appropriate technique for calculating geometric entropy in
conformal field theory.  In Section 3 we apply this machinery to the
moving mirror model, showing how the geometric entropy arises naturally
in a dynamical context, how ultraviolet divergences arise and how a
useful renormalized entropy can be defined.  Finally in Section 4 we
briefly review the use of the moving mirror
model in black hole physics, and attempt to draw lessons from our work
regarding the corresponding problems of black hole entropy.

\chapter{Geometric Entropy in Conformal Field Theory}

\section{Geometric Entropy in General}

In statistical mechanics one does not resolve, but rather
averages over,
physically distinct states of a system
which have common values of macroscopic state
variables.
Many
microscopically different states look alike macroscopically.
Entropy is a precise measure of this lack of resolution; roughly speaking,
the entropy of a macroscopic state
is the logarithm of the number of microscopic states with which it is
consistent.

In quantum physics there is an additional,
conceptually distinct source of entropy
associated with the limitation of experiments to a finite volume.
Even if the universe as a whole -- or an idealized ``closed system'' --
is taken to be in a definite pure
state, say the ground state, a complete description of
the information available
to an observer who has access only to a partial set of the
observables, such as those with support in a restricted volume, will
be given by a non-trivial density matrix $\rho$.  It is natural to
use the same definition of entropy in this circumstance as one uses for
density matrices associated with averaging over macroscopically equivalent
microscopic states in statistical mechanics, that is,
$$
 S = - {\rm tr} \rho {\rm ln} \rho~.
\eqn\sa
$$
This entropy describes correlations
between the subsystem and the rest of the universe.
Roughly speaking, it the logarithm of the number of states
of the inaccessible part of the
universe that are consistent with
all measurements restricted to the accessible part, together
with {\it a priori\/} knowledge that the universe as a whole is in a
pure state.

In quantum field theory the principle of
locality is embodied directly,
and there is a particularly natural, precise
concept of {\it geometric entropy\/} along these
lines [\bombelli --\susskindfour ].
Explicitly, the
geometric entropy of a region $R$ relative to a state (pure or mixed) $U$
of the universe is defined as follows.
Define a complete set of commuting observables
$\hat \xi_{\rm in}, \hat \xi_{\rm out}$
that are  localized respectively
completely within and completely outside $R$.  The density matrix
of the universe $\rho_{\rm U}$ can be expressed as a function of
the eigenvalues of these variables; to wit
$$
\rho_{\rm U} ~=~
\rho_{\rm U} (\xi^1_{\rm in}, \xi^1_{\rm out}~;~
\xi^2_{\rm in}, \xi^2_{\rm out})~.
\eqn\dfnofrhoU
$$
Then the density matrix for observations restricted to the inside
is
$$
\rho_{\rm in} (\xi^1_{\rm in}~;~\xi^2_{\rm in} )~=~
\Sigma_{\xi_{\rm out}}~
\rho_{\rm U} (\xi^1_{\rm in}, \xi_{\rm out}~;~
\xi^2_{\rm in}, \xi_{\rm out})~.
\eqn\dfnofrhoin
$$
In a simple scalar field theory one could use the
field operators $\hat \phi (x)$ (at some definite fixed time)
for the required
set of observables.  The heart of the matter is that these operators,
being
defined as local functions of position, are in an obvious way either
inside or outside a specified region.

Thus far our discussion has been purely abstract and formal.  It is
notorious that divergences can occur for formally defined
quantities
in relativistic quantum field theories.  The most fundamental
divergences of this kind, with which we shall mainly be concerned below,
arise from the singular
ultraviolet or short-distance behavior of the theories.
In our present context of thermodynamics and information theory, it is
perhaps most suggestive to say that they arise from the existence of
an infinite number of degrees of freedom per unit volume.
In favorable cases one knows
how to regulate and renormalize, in such a way that
physically meaningful quantities
are assigned definite finite values in the theory.  It is perhaps not
obvious that geometric entropy as defined above is a directly physically
meaningful quantity, that must be finite in any realistic
theory.  For example
no realistic measuring apparatus resolves infinitely
small distances, so the sharp distinction between inside and outside
might appear to be an unrealistic
idealization.  As we shall see, a sharp cut-off typically
brings in divergent entropy
from the singular short-distance behavior of
relativistic quantum field theory, which
implies strong correlations between observables
near the boundary.
Nevertheless we shall argue below that geometric entropy arises very
naturally in interesting physical problems, so that its divergence is
of interest in and of itself.
Considering how the regularized geometric entropy of a region
varies as a function of the state of the whole universe, we define finite
renormalized
relative entropies, that appear to be quite meaningful physically.

\section{Evaluations in Conformal Field Theory}

To make the discussion concrete and explicit,
we now specialize to the case where the field theory
in question is a conformal field theory in (1+1) dimensions. Furthermore,
in this section
we consider only the vacuum state of the theory.

Introducing
an infrared cutoff $\Lambda$,
we take our universe to be ${\cal C}=[0,\Lambda [$ with
periodic boundary conditions defining the region outside $\cal C$.
The subsystem where measurements are performed is ${\cal R}_1=[0,\Sigma [$.
The degrees of freedom in the region ${\cal R}_2=[\Sigma,\Lambda [$ are to
be traced over.
Now, the entropy \sa~turns out to be
infinite, because the problem as defined so far has
no ultraviolet cutoff.  Therefore localized excitations arbitrarily
near the boundaries of the subsystem can correlate the subsystem
with the rest of the universe, and they contribute arbitrarily much
to the entropy. To regulate this, we introduce a smearing at
the ends of the subsystem.  Specifically, we take the ends to be at
$\pm\epsilon_1$ and at $\Sigma\pm\epsilon_2$, instead of at $0$ and at
$\Sigma$. Here
$\epsilon_i, i=1,2$ are coarse graining
parameters that parameterize how well the observer distinguishes
the subsystem from the rest of the universe.
As we shall now show,
the microscopic entropy grows as $\epsilon_i$ becomes smaller
and it diverges as $\epsilon_i\rightarrow 0$.
We will show that the divergence is logarithmic and calculate its
coefficient.

Conformal field theories, of course,
respond in a simple way to
conformal transformations.
Such transformations are implemented as unitary
transformations, and
the vacuum is
invariant under global conformal transformations.
(The representation of conformal
symmetry is projective, so that strictly speaking there
is no way to choose the phase of the various transforms
of the vacuum in a globally consistent way.  The
projective nature of the representation is variously
manifested in the
existence of a central charge in the Virasoro algebra, the
anomalous transformation law of the energy-momentum tensor, and
the trace anomaly.   This subtlety,
though it does not affect the present argument, will play
an important role both implicitly and explicitly
below. ) From this we readily conclude that the geometric entropy
relative to the vacuum is
invariant.  Indeed a conformal change of coordinates
simply induces a change of basis (unitary transformation)
among the
operators
of the theory, without altering their character
as inside or outside nor their spectrum, and the trace \sa~is manifestly
invariant.

Thus we
can use conformal mappings to simplify our calculations.
This potential is most easily exploited by
introducing complex coordinates, as follows. Let
$\zeta=\sigma+i\tau$, where $\sigma$ is the spatial coordinate and
$\tau$ is the time coordinate, with $\tau=0$ defining
the Cauchy surface $\cal C$.
We first make the problem more symmetric and canonical by
mapping
$$
w = -{ {\rm sin}{\pi\over\Lambda}(\zeta-\Sigma)\over {\rm sin}
{\pi\over\Lambda}\zeta }
\eqn\conftoaxis
$$
This transformation maps
the subsystem to the positive half--axis and the
rest of the universe to the negative half--axis.
In the limit where $\Sigma\ll\Lambda$ the cutoffs
are mapped to $\pm{\epsilon_2\over\Sigma}$ and
$\pm{\Sigma\over\epsilon_1}$.  The infrared
cutoff $\Lambda$ decouples from the ultraviolet
cutoffs in this limit, allowing a clean separation
to be made.
We extrapolate off the
real axis by picking our system as an annulus restricted to
the lower half--plane, having inner and outer radii $\epsilon_2\over\Sigma$
and $\Sigma\over\epsilon_1$, respectively. This choice amounts to a
convenient specification of how the smearings of the endpoints of the
subsystem are extrapolated from $\tau=0$
into the past.  Details of this extrapolation are unimportant
due to conformal invariance. The only function of the extrapolation
to the past is that imposing
regularity there selects
the vacuum state.

Now we transform
$$
z = {1\over\kappa} {\rm ln} w
\eqn\logtransform
$$
where $\kappa$ is an auxiliary parameter.  $\kappa$ has no
independent physical
meaning, and hence it must not appear appear in the
final result.
The transformation \logtransform\ maps
our system on to a finite strip of width $\pi\over\kappa$
and length $L= {2\over\kappa}{\rm log}{\Sigma\over\epsilon}$,
where for simplicity
we have chosen a symmetric cutoff $\epsilon=\epsilon_1=\epsilon_2$.
(One can recover the general case by
substituting $\epsilon=\sqrt{\epsilon_1\epsilon_2}$.)

The interval ${\cal R}_1$ representing the accessible subsystem
is now the upper side
of the strip, and the interval ${\cal R}_2$ representing the
rest of the universe is the lower side of the strip.
We impose periodic
boundary conditions in the length direction of the strip.
This amounts to a specification
of the fields within the smearing intervals at the ends of the original
subsystem.  We shall argue below that
the details of how these fields are specified
within the smearing intervals does not affect our main results.
Our sequence of mappings is shown in Figure 1.

With the specified regulators in place,
the wavefunction of our system is
$$
\Psi_{XY} \propto \int {\cal D}\phi e^{-S(\phi)}.
\eqn\psia
$$
where $\phi$ denotes a complete
collection of local fields on our theory.
In the functional integral boundary conditions
specify the fields on the Cauchy surface
${\cal C}={\cal R}_1\cup{\cal R}_2$, where ${\cal R}_1$ and
${\cal R}_2$ is the upper and lower part of the strip, respectively.
We take $\phi=X$ on ${\cal R}_1$ and $\phi=Y$ on ${\cal R}_2$,
where $X$ and $Y$ are ordinary c-number functions\foot{A number
of subtleties arise for fermions.  They will be discussed in
a separate paper.}. The density
matrix describing the subsystem on ${\cal R}_1$ after tracing over
variables on ${\cal R}_2$ is
$$
\rho_{XX^{\prime}}=\int {\cal D} Y \Psi_{XY}\Psi_{YX^{\prime}}^{\ast}
\eqn\rhoa
$$
In general, after the integration
$\rho$ can no longer be written in the factorized form
$\rho_{XX^{\prime}}=a_X a_{X^{\prime}}$ for any function $a_X$.
Thus the system is in a mixed state.

Indeed, in the present coordinates it is described by a
very specific and familiar mixed state.  For the integral in
\rhoa\ can be represented by pasting together two copies of the
strip along ${\cal R}_2$.  Then we have the fields specified on two
sides of a strip of width ${2\pi \over \kappa}$.  Interpreting
the
functional integral as implementing evolution in imaginary time, one
realizes that
we have specified
the matrix element of the {\it thermal\/} density matrix at
inverse temperature $\beta={2\pi \over \kappa}$.

For free field theory one could easily
characterize the functions $X$ and
$Y$ by their Fourier components, and carry out the evaluation
of the wave functional and the density matrix
explicitly.  One can also make progress in more general cases.
However for our
present purpose of evaluating the entropy
it will be sufficient to work with the foregoing abstract
expressions, valid for any conformal field theory
described by a Lagrangean.

Inserting \psia~in \rhoa~ and normalizing we find
$$
\rho_{XX^{\prime}}= {1\over Z(1)}
\int {\cal D}\phi e^{-S(\phi)}~.
\eqn\rhob
$$
Here the functional integral is over a strip of height $2\pi\over\kappa$
with boundary conditions $\phi=X$ on the upper side and $\phi=X^{\prime}$
on the lower side. $Z(1)$ is determined by the condition
that ${\rm tr}\rho =1$, so it is given by the
same functional integral expression but with periodic boundary
conditions on top and bottom.  Since we have already imposed
periodic boundary conditions in the length direction of
the system, $Z(1)$ is the partition function on a torus.

The entropy corresponding to the density matrix \rhob~
is calculated using the replica trick
$$
S = - {\rm tr} \rho{\rm ln}\rho =
- ({d\over dn})_{n=1}{\rm tr}\rho^{n}
$$
Here $\rho^n$ is found by first calculating $\rho^n$ for
integers $n$ and then analytically continued to general $n$.
In our case \rhob~ gives
$$
\rho^n_{XX^{\prime}}= {1\over Z(1)^n} \int {\cal D}\phi e^{-S(\phi)}
$$
where the integral is over a strip with width ${2\pi n}\over\kappa$.
This is trivially extended to general $n$. Taking the trace we find
$$
S= - ({d\over dn})_{n=1}
{Z(n)\over Z(1)^n}= (1-n
{d\over dn})_{n=1}{\rm ln}Z(n)~.
\eqn\zn
$$
Here the symbol $Z(n)$ denotes the
partition function on a torus which measures $2\pi n/\kappa$
and $L$ around the two cycles.

Let us illustrate this formalism by the case of
free scalar (massless) bosons. The
partition function is readily calculated
by expanding in normal modes. Choosing periodic boundary
conditions, for definiteness, one finds
$$
Z(n)= {1\over\eta\bar\eta}
\eqn\ztorus
$$
where
$$
\eta = q^{1\over 24}\prod^{\infty}_{k=1}(1-q^k);~~~q=e^{2\pi i\tau}
$$
in terms of the modular parameter $\tau=i {2\pi n\over\kappa L}$.
Except for the factor $(q\bar q)^{-1\over 24}$, which we have included
for a reason to be discussed presently,
the partition function has the same form as that
of that for
free photon gas with inverse temperature $n$ and energy levels
$k$ in
appropriate units, as anticipated.
Noting that ${\rm ln}q\propto n$, which is proportional to the
height of the strip, i.e. the inverse temperature,  we see that
in the expression
$$
S = (1-{\rm ln}q{\partial\over\partial {\rm ln}q}
-{\rm ln}\bar q{\partial\over\partial {\rm ln}\bar q})
{\rm ln}Z (1)
\eqn\sc
$$
the contribution from the extra prefactor cancels, so that
the entropy has the
the standard thermodynamic form $S=\beta(F-E)$ in appropriate units
\foot{We are adhering to the convention that $q$ and $\bar q$ are to
be treated as independent variables.  They describe the contribution of
holomorphic and anti-holomorphic, or left- and right-moving, modes,
respectively.  Since we are treating them symmetrically, we are implicitly
dealing with a non-chiral scalar field.  For a free
chiral scalar, the entropy
is of course just half as large.}.
The calculation of geometric entropy therefore
reduces to the elementary calculation of the entropy of a thermal photon
gas in one dimension.  One thereby finds
$$
S = {1\over 3}{\rm ln}{\Sigma\over\epsilon}~.
\eqn\zaza
$$
We note with relief that the parameter $\kappa$ of the
coordinate transformation does not appear in the final result.  It cancels
between the entropy per unit volume and the volume.
The entropy is divergent as $\epsilon\rightarrow 0$, as advertised.

Now let us return to discuss the prefactor.  It arises if we
write the unnormalized density matrix as
$\tilde \rho = \langle e^{-\beta H} \rangle$
and take into account the anomalous transformation law of the
energy-momentum tensor.  The additional c-number contribution to the
Hamiltonian resulting from this transformation law resets the zero of
energy.  Correspondingly it changes the normalization of $\tilde \rho$,
and the partition function, but not the normalized density matrix nor
the geometric entropy, consistent with our previous argument.

In the preceding
discussion we have assumed, for convenience, that $\Sigma\ll
\Lambda$.  One can go through the same manipulations without making this
assumption; the result is
$$
S~=~{1\over 3}\ln \bigl({\Lambda \over \epsilon}
\sin {({\pi\Sigma \over \Lambda}) \over \pi} \bigr)~.
\eqn\intent
$$
The entropy is the same
for inside and outside ($\Sigma$ and
$\Lambda - \Sigma$), as it should be.
In using this expression we must still assume $\Lambda - \Sigma \gg
\epsilon$ and $\Sigma \gg \epsilon$.  With this restriction, note
that the entropy reaches a maximum when
$\Sigma = \Lambda/2$, and then decreases as $\Sigma$ increases
further.  This is as it should be -- when the subsystem begins to fill
most of the universe, there is less information to be lost by not
measuring outside.
As $\Lambda-\Sigma$ or $\Sigma$ becomes comparable to $\epsilon$ the
entropy becomes of ${\cal O} ( 1 )$, and it looses physical meaning as
it becomes dependent on the details of the regularization.

Srednicki [\srednickione ,\srednickitwo ]
has reported numerical results for the geometric entropy.
He also finds a logarithmic divergence and the coefficient agrees
approximately with \zaza\ . However, he also finds an additional term
that depends on the infrared cutoff $\Lambda $. Some of this dependence
may be described by \intent\ , but we should also point out
that our calculation has ignored 0-modes.
If these are allowed by the boundary conditions,
it can be argued that they add a contribution that depends on the
infrared cutoff, [\srednickitwo --\keski ].

The entropy may be found in a different way,
that has its own
intrinsic interest and does not appeal to
ordinary thermodynamics as a {\it deus ex machina}.
The partition function \ztorus\
is invariant under the modular transformation
$\tau\rightarrow -{1\over\tau}$, so
we can take $\tau={i\kappa L\over 2\pi n}$.
Now
${\rm ln} q\propto 1/n$ so \zn~ gives
$$
S = (1+{\rm ln}q{\partial\over\partial{\rm ln}q}
+{\rm ln}\bar{q}{\partial\over\partial{\rm ln}\bar{q}}
){\rm ln}Z(1)
\eqn\sd
$$
This differs
from \sc~by the sign of the logarithms. We find
$$
S = - {1\over 6}{\rm ln}q -2(1+{\rm ln}q{\partial\over\partial{\rm ln}q})
\sum_{k=1}^{\infty} {\rm ln} (1-q^k)
$$
But $q=e^{-\kappa L}$ is exponentially small, so we can omit the last
term and recover
$$
S=-{1\over 6}{\rm ln} q={1\over 3}{\rm ln}{\Sigma\over\epsilon}.
$$
The modular transformation vastly simplified the calculation:
we started with a system which had an excitation spectrum like
the free photon gas, and to calculate the entropy we needed to
carry out the sum over all those many--particle degrees of freedom.
In contrast,
after the modular transformation only a seemingly inoccuous vacuum
piece contributed. The presence or absence of
the prefactor $(q\bar{q})^{-1\over 24}$ did not affect
the original calculation since it cancels in \sc\ .
It does, however, ensure invariance under the modular transformation,
after which the prefactor contained all the information needed.
Geometrically the
transformation $\tau\rightarrow -{1\over\tau}$
amounts to interchange of width and length of the torus,
a transformation that is simply a change in bookkeeping.
However, it is well known in conformal field theory
and string theory that in the Hamiltonian interpretation
of the partition function,
the modular transformation relates the many highly excited states to
the few low lying states in a non--trivial way.  That is why,
in our context,
the contribution from all the many--particle states can be
found with so little effort.

The latter calculation of the entropy can be generalized immediately to
the general conformal field theory. We still have formula \sd~for
the entropy. For a conformal field theory
with central extensions $c$ and $\bar{c}$,
the partition function on the torus is
[\ginsparg ]
$$
Z(\tau,\bar\tau)= q^{-{c\over 24}}\bar{q}^{-{\bar{c}\over 24}}{\rm tr}
q^{L_0}\bar q^{\bar L_0}
\eqn\general
$$
with $q=\bar q=e^{-\kappa L}$. So
$$
{\rm ln}Z=-{c+\bar{c}\over 24}{\rm ln}q + {\rm ln}{\rm tr} q^{L_0+\bar L_0}
\eqn\za
$$
Now we expand
$$
{\rm tr}q^{L_0+\bar L_0} = 1 + q^{\alpha}+ ...
\eqn\expansion
$$
where $\alpha>0$ and the
dots denote yet higher positive powers of $q$.
The existence of an expansion in positive powers of $q$ is due to the
requirement of positive dimensions of all fields in the theory,
which is needed to ensure locality. With this expression we see that
the last term in \za~is exponentially suppressed and we find
$$
S={c+\bar{c}\over 6}{\rm ln}{\Sigma\over\epsilon}
\eqn\se
$$
for the general conformal field theory.
The derivation shows that this expression is {\it exact}, i.e.
corrections corresponding to higher powers of $q$
vanish as $\epsilon\rightarrow 0$, rather than just being subleading.

We should emphasize that the general calculation does not assume
that the partition function on the torus is invariant under the
full group of modular transformations. In fact, we merely use that
after the transformation $\tau\rightarrow -{1\over\tau}$ the partition
function has the form \general\ with an expansion \expansion\ .
In our interpretation of the partition function,
one of the cycles of the torus corresponds to taking the trace of
$\rho^n$. In this direction physics demands periodic boundary conditions
(for bosonic variables). In contrast extra twists along the other cycle
do have a legitimate interpretation as a change of spatial boundary
conditions. The modular transformation changes the cycles so as to
forbid twists along the spatial cycle instead. Hence
twists cannot contribute to the vacuum energy of the Hamiltonian, i.e.
$L_0 + \bar{L}_0$ has 0 as eigenvalue in the vacuum state. This is
exactly what we use.

There is a caveat: the discussion so far assumes that the
vacuum is non-degenerate. This is amounts to ignoring 0-modes.
Including them can change the normalization of the partition function
by a power of the modular parameter $\tau$. This may introduce additional,
but subleading, dependence on the ultraviolet cutoff.
We will continue to ignore the 0-modes.

\section{Another Approach}

The preceding derivations of \se~ do not bring out
the physical origin of the
logarithmic divergence with optimal clarity.
We will now therefore rederive \se~ in a manner which stresses
how the entropy arises through coarse graining
in real space, without explicit use of mode
expansions or modular invariance.
The calculation follows ideas of Cardy [\cardy ].

We use the $w$ coordinates, where the system is represented by
an annulus restricted to the lower halfplane.
The positive real axis is the subsystem where observations are
made, and the negative real axis is the rest of the universe.
Forming the density
matrix of the accessible subsystem we trace over variables on the negative
real axis. Then the density matrix is given by a functional integral over
fields over the whole annulus, with the indices $X$ and $X^{\prime}$
of $\rho_{XX^{\prime}}$ specifying fields on
the lower and upper side of the positive real axis.
As before we use the replica trick
$$
S=(1-n{d\over dn})_{n=1}{\rm ln}Z(n)
$$
where $Z(n)\propto{\rm tr}\rho^n$ is the partition function
of an annulus covered $n$ times.
Extending $n$ analytically to be slightly
less than $1$, we can interpret $Z(n)$ as the partition
function on a cone with $2\pi n$ in angular circumference.

Now we coarse grain the system by taking
$\epsilon\rightarrow (1+\alpha)\epsilon$.
The annulus has outer radius $R_2={\Sigma\over\epsilon}$ which decreases
and inner radius $R_1={\epsilon\over\Sigma}$ which increases. By dimensional
analysis ${\rm ln}Z$ depends only on the ratio of the two radii,
so we can choose to rescale only the outer radius twice as much and keep
the inner one fixed. This is implemented by the rescaling
$x^{\mu}\rightarrow x^{\prime\mu}=(1-2\alpha)x^{\mu}$, in the
limit where not only $\alpha$ but also $R_1$ is treated as small,
i.e. we squeeze the inner boundary to a conical singularity.
Using conformal invariance of the path integral measure
and the definition of the energy momentum tensor
as the generator of coordinate transformations it is easy to show
$$
\delta{\rm ln}Z = -{1\over 2\pi}\int \langle T_{\mu}^{\nu}\rangle
{\partial x^{\prime\mu}\over\partial x^{\nu}}d^2 r
= {\alpha\over \pi}\int \langle T_{\mu}^{\mu}\rangle d^2 r
$$
With $\alpha={\delta\epsilon\over\epsilon}$ we find
$$
{\partial S\over\partial{\rm ln}\epsilon}
=(1-n{d\over dn})_{n=1}{\partial{\rm ln}Z(n)\over\partial{\rm ln}\epsilon}
=(1-n{d\over dn})_{n=1}{1\over \pi}\int\langle T_{\mu}^{\mu}\rangle d^2 r
\eqn\sef
$$
The trace of the energy momentum tensor is related to
the curvature of the manifold and its boundaries [\cardy ].
Inserting the appropriate expressions \se~ can be recovered.

We find it illuminating to proceed
slightly differently, writing
$$
\int \langle T_{\mu}^{\mu}\rangle d^2 r =
\int {\partial x^{\mu}\over\partial x^{\nu}}
\langle T_{\mu}^{\nu}\rangle d^2 r
= \int x^{\mu}\langle T_{\mu\nu}\rangle dS^{\nu}
= -i \int w\langle T(w)\rangle dw + {\rm h.c.}
\eqn\trt
$$
where we introduced complex coordinates. Performing the surface integral
we are to integrate over the outer surface only.
The expectation value of $T(w)$ on a cone with angular circumference
$2\pi n$ is easily found by mapping to the cone with $n=1$ which
is simply a disc.
We map $w=y^n$ and impose $\langle T(y)\rangle=0$ on the disc.
This is appropriate to our problem, since the primary object of study
is the geometric entropy relative to
the ground state on the disc.
Using the standard transformation formula for a conformal field theory
with central charge $c$,
$$
T(w)=({\partial y\over\partial w})^2 T(f(y)) + {c\over 12}S_y(w);
{}~~S_y(w)={y^{\prime\prime\prime}y^{\prime}-
{3\over 2}(y^{\prime\prime})^2\over (y^{\prime})^2}
\eqn\ttrans
$$
we find
$$
\langle T(w)\rangle={c\over 24} (1-{1\over n^2}) {1\over w^2}~.
$$
Inserting this in \trt~, we find
$$
\int \langle T_{\mu}^{\mu} \rangle d^2 r=
{c+\bar{c}\over 24} (1-{1\over n^2})2\pi n
$$
which we insert in \sef~ to
find
$$
{\partial S\over\partial{\rm ln}\epsilon}= - {c+\bar{c}\over 6}
$$
By integrating this
we recover \se~. Notice that
in this procedure a finite $\epsilon$--independent term
cannot be excluded.

In this derivation the conformal anomaly plays a crucial role.
Formally, and classically,
the trace of the energy-momentum tensor vanishes as a consequence
of conformal invariance; but the necessity of regulating the
quantum theory brings in the correction term $c$.
This term corresponds directly
to extra correlations in the products of energy-momentum tensors at
short distances, appearing in the operator product
$$
T(z)T(z^{\prime} ) ~\rightarrow {c\over  2}  (z - z^{\prime})^{-4}
{}~+~ {\rm less\ singular}~.
\eqn\opprod
$$
Thus the divergence in the geometric
entropy can be traced, quite directly, to the
singular short-distance behavior of quantum field theory.

\chapter{Renormalized Entropy for Moving Mirror States}

In the previous Section we have discussed the concept of geometric
entropy, and evaluated it for finite
intervals relative to the ground
state of conformal field theories in 1+1 dimensions.
We found that it diverges in the absence of an ultraviolet cutoff, and
must be regulated.  Since the high-energy modes responsible for the
divergence are not easily excited, however, we might expect that
the divergent piece of the geometric entropy will not change if we evaluate
it relative to some other low-energy state.  This suggests that the
{\it difference\/} between the geometric entropy of a given state and
that of the vacuum is a finite quantity characterizing an interesting
physical property of the state.

Now we proceed to a simple class of excited states for
which the geometric entropy
is readily evaluated (and, as we shall see, has an interesting
physical interpretation.)
We consider
a conformal field theory in 1+1 dimensions
with a boundary at ${\bar z}=f(z)$, where $f(z)$ is
an arbitrary function.  We require
the fields in the theory to
vanish on the boundary, which can therefore be interpreted
as a
perfectly reflecting moving mirror.
For definiteness we choose to consider the model
to the right of the boundary.

To an observer far to the right of the mirror,
the moving mirror manifests itself as a change in the radiation
field compared to the case of a stationary mirror.  Thus each mirror
trajectory corresponds to a state. It is natural to identify
the stationary mirror with the vacuum.

The mirror model is  easily solved by performing the
conformal transformation $z\rightarrow f(z)$. This relates
the moving mirror to a stationary mirror, which is trivial.

Let us use this procedure to
calculate the microscopic entropy as seen by a
distant observer. As Cauchy surface we choose a line of
constant $\bar z$. This surface is light-like, so the material
observer can not choose it as his or her world line; nevertheless
it is
possible to monitor an interval $[z_1,z_2]$ on the Cauchy
surface by appropriate organization of the measuring apparatus.
For right-moving modes, as we consider, it is only necessary to
monitor a surface that intersected by the same light-rays; this surface
can be chosen space-like or even at a fixed time. Applying \se\ we should
take $\bar{c}=0$ since we only consider one set of modes.

It is natural for the observer to choose the smearing
at the ends of the interval symmetrically as seen in his or her
coordinate system. Since
$$
\epsilon_f=f^{\prime}(z)\epsilon_z
$$
this choice corresponds to an asymmetric choice in the $f$--coordinate
system, where the mirror is stationary. This apparently technical point
makes a world of difference, as we shall now see.
Expression \se~ for the entropy in the vacuum state is valid in the
coordinate system where the mirror is stationary. Recalling
$\epsilon=\sqrt{\epsilon_1\epsilon_2}$ for an asymmetric
choice of smearing, we find
$$
S_{\rm bare}={c\over 6}{\rm ln}{\Sigma\over\epsilon}={c\over 12}{\rm ln}
{\Sigma^2\over\epsilon_{f,1}\epsilon_{f,2}}={c\over 12}{\rm ln}
{(f(z_2)-f(z_1))^2\over f^{\prime}(z_1)f^{\prime}(z_2)\epsilon^2_z}
$$
Clearly the entropy of the system is infinite in the limit
$\epsilon_z\rightarrow 0$.

However, the observer would find this infinity
even if the mirror were not moving at all, {\it i.e}. if
observation were made in vacuum.
It is therefore natural to define
$$
S_{\rm ren}=S_{\rm bare}-S_{\rm vac}
$$
where $S_{\rm vac}$ is the entropy expected for a stationary mirror,
that is, for $f(z)=z$.
The renormalized entropy $S_{\rm ren}$ is
$$
S_{\rm ren}={c\over 12}{\rm ln}
{(f(z_2)-f(z_1))^2\over (z_2-z_1)^2 f^{\prime}(z_2)f^{\prime}(z_1)}
\eqn\sr
$$
$S_{\rm ren}$  is independent of the smearing $\epsilon_z$, and
in particular it is finite as $\epsilon_z\rightarrow 0$.
This is the physical entropy. It is a property of the state of
the system, which expresses the information content of the state.

\sr\ is a central result of our analysis.   As we shall see in the
following
Section, it has an interesting application to black hole physics.
Before discussing that, however, we would like to elucidate the
meaning of \sr\ by eliminating the function $f$ in favor of physical
variables directly accessible to our observer.

Consider the
energy--momentum tensor $T(z)$. We require
$\langle T(f(z))\rangle =0$ for the stationary
mirror, and again invoke the transformation law
$$
T(z)=f^{\prime}(z)^2 T(f(z)) + {c\over 12}S_f(z);
{}~~S_f(z)={f^{\prime\prime\prime}f^{\prime}-
{3\over 2}(f^{\prime\prime})^2\over (f^{\prime})^2}~.
\eqn\ttrans
$$
Thus we find a non--zero $\langle T(z)\rangle $, unless
$S_f(z)$ happens to vanish.  A non-vanishing result
corresponds to a flux
of particles away from the mirror. When the moving mirror
is interpreted as a model for a black hole, this is
the Hawking radiation.
The 2--point correlations are
$$
C(z_2,z_1)\equiv \langle T(z_2)T(z_1)\rangle  -\langle T(z_2)\rangle
\langle T(z_1)\rangle =
{c\over 2}
{f^{\prime}(z_2)^2f^{\prime}(z_1)^2\over (f(z_2)-f(z_1))^4}
$$
where we have used the standard result
$$
\langle T(f(z_2))T(f(z_1))\rangle ={c/2\over (f(z_2)-f(z_1))^4}
$$
for the plane, which is also valid for the half--plane.
The function $C(z_2,z_1)$ describes correlations in the energy
momentum observed.

For the trajectory
$$
f(z)~=~ c_1 + c_2e^{-z/4M}
\eqn\bh
$$
which arises when the moving mirror is used as a model for
Schwarzschild geometry,
the correlation function is thermal with temperature $T_H={1\over 8\pi M}$.
However not all of these correlations can be attributed to
the Hawking radiation, since even for a stationary mirror we expect
correlations, namely the vacuum correlations.

In any case, we can describe the
excess of correlations relative to vacuum by dividing the
correlation function $C(z_2,z_1)$ by its value $C_0(z_2,z_1)$
expected for a stationary mirror.  Then an
intriguing coincidence emerges:
$$
S_{\rm ren} = -{1\over 24}{\rm ln}{C\over C_0}~.
\eqn\scorl
$$
\scorl\ allows a heuristic interpretation of the
renormalized entropy, as a measure of correlations in
the observed energy--momentum tensor being in excess of
the correlations expected in vacuum.  Notice that
if there are more correlations
in the observed radiation than expected in vacuum $S_{\rm ren}$ is
negative! This is as it should be, because it corresponds to the
state being more ordered than vacuum, which has vanishing entropy.
The locally negative entropy found here is very reminiscent of the
locally negative energy which emerges in analyses of the Casimir effect
and of vacuum polarization near black hole horizons.
\scorl\ is physically very reasonable, and it links
the renormalized entropy to correlations in
as concrete a manner as
one could desire.  Whether it has a useful generalization outside the
immediate context of the moving mirror problem, is a question worthy
of further investigation.

\chapter{Application to Black Holes}

\def\se{S_{exp}}
\def\sa{S_{act}}

\overfullrule=0pt

\def\Gv{G_{vac}(1,2)}

\section{Moving Mirrors, Collapse, and Radiance}

We will now briefly review the well-known connection between
the mirror model and collapse geometry, in a language consistent
with our previous discussion. We use a notation that is conventional
in black hole physics [\birrell]. It differs from the conformal field theory
notation used so far.

Consider for simplicity
a spherically symmetric collapsing shell of matter.
We have
vacuum inside and outside the shell, while the shell carries a given amount
of mass (and possibly other quantum numbers).
Thanks to Birkhoff's theorem, we know the
metric in both  regions:
$$ds^2= \cases{dr^2-d \tau^2-r^2 d\Omega^2,
                & for $\tau+r \le V_s $;\cr
        \lambda^2 dt^2-\lambda^{-2} dr^2 -r^2 d \Omega^2,
                & for $t+r \ge v_s$. } \eqn\meta$$
Note that in order to exhibit the metric in each region in its familiar
(static) form, two different sets of coordinates had to be used.   It is
convenient to introduce light-cone coordinates in each region. In
the interior region we use simply
$U=\tau-r$ and $V=\tau+r$, whereas in the outer region we first define the
tortoise-coordinate $r_*$ through
 $${dr_* \over dr}= {1 \over \lambda^2}, \eqn\rtortdef $$  and then take
$u=t-r_*$ and $v=t+r_*$ as light-cone coordinates. The  space-time is
described by the metric:
$$ds^2=\cases{dUdV-r^2 d\Omega^2, & for $V\le V_s$; \cr
        \lambda^2 dudv-r^2 d\Omega^2, & for $v\ge v_s$ ,}
        \eqn\metb$$
where $r$ is determined through the relations
$$\eqalign{V-U=2 r, \qquad & {\rm for }\quad  V\le V_s; \cr
                v-u=2 r_*(r), \qquad & {\rm for }\quad  v\ge v_s.}
         \eqn\rofuv $$

When we paste together the two coordinate systems for the interior and
exterior region to form a global coordinate-system, we can choose to
coincide with \metb\ either
in the exterior or in the interior region. The first choice
is natural from the point of view of a distant observer,  while the second
is more convenient to implement the boundary condition at the
origin and to display the complete space-time structure.

Let us consider first the
former choice, that is
using $u$-$v$-coordinates in both regions and looking for a
satisfactory coordinate-transformation $U(u)$
and $V(v)$. In the infinite past the space-time is flat and there is no
difference between the two coordinate systems. This implies that we can
choose $V(v)=v$. We find the function $U(u)$ by demanding that along
the worldline $v=v_s$ of the shell the coordinate $r$ should agree in both
systems, because it has a coordinate invariant meaning (it determines the area
of a two-sphere at constant radius and time). Applying \rofuv\  along
$v=v_s$ we  obtain the  implicit relation:
$$r_*\left(r={v_s-U(u)\over 2} \right)= {v_s-u \over 2}. \eqn\Uofu $$
Differentiating  this equation along the worldline of the shell we find,
with the help of the defining equation \rtortdef\ for $r_*$,
$${dU \over du}=\lambda^2(u,v_s),
	\eqn\dUdu$$
so that the metric becomes:
$$ds^2=\cases{\lambda^2(u,v_s) dudv-r^2 d\Omega^2,
         & for $v<v_s$; \cr
        \lambda^2(u,v) dudv-r^2 d\Omega^2, & for $v>v_s,$ }
        \eqn\metuv$$
which is continuous along $v_s$. The metric is, of course, only valid for
non-negative values of $r$, \ie\ for $v \ge U(u)$. The world-line of the
origin is therefore described by
$$v_o(u)=U(u). \eqn\origineq $$
Since nothing can go beyond the regular origin, \ie\ to negative $r$, it
acts like a perfectly reflecting mirror.

In the $u$-$v$-frame the shell never crosses the horizon since $r_*$
and $t=v_s-r_*$ diverge as the horizon is approached. On the other
hand we know that the shell reaches the origin in finite proper time.  In
order to describe the whole space-time, including the interior of the black
hole
it is convenient to use the
$U$-$V$-coordinates, which provide a complete cover since they contain
the  origin until the shell reaches it. The space-time is then described by
$$ds^2=\cases{ dUdV-r^2 d\Omega^2,
         & for $v\le v_s$; \cr
        \lambda^2(u,v) \lambda^{-2}(u,v_s) dUdV-r^2 d\Omega^2,
         & for $v\ge v_s,$}
        \eqn\metUV$$
In spite of its appearance, the metric is regular on the horizon where
$\lambda^2=0$. The origin is stationary at $V=U$ until the shell reaches it.

For a shell of mass $M$ one has explicitly for
the tortoise coordinate
$$r_*(r)=r+2 M \ln |r-2 M|+c,
	\eqn\rtorts$$
and thus from \Uofu
$$u=U-4M \ln\left|(-4M-U+v_s)/2\right|-2 c. \eqn\uschwarz $$
$c$ is here an arbitrary integration constant. As $U$ approaches
$U_h=v_s-4 M$, $u$ diverges, which identifies  the line $U=U_h$
with the future horizon.  Alternatively, the finite range of $U$ implies
according to \origineq\ that the origin approaches the light-like asymptote
$v=U_h$ at late times as viewed in the $u$-$v$-frame. At very early times,
the origin is at rest because as $u\to -\infty$,
$U \approx u$.  At late times we can invert
\uschwarz\  by neglecting the linear term. We  find that $U(u)$
is of the general form
$$U(u) = c_1+c_2 e^{- \kappa u},  \eqn\trath$$
where $\kappa=1/4M$ is the surface gravity.  This relation is nothing but
the familiar transformation between Eddington-Finkelstein
and Kruskal-coordinates:
$$U_K =-4 M e^{-u/4M}.
	\eqn\Ukrus$$
At late times our coordinate $U$ therefore agrees with Kruskal $U_K$,
while $V$ equals $v$  is always of the Eddington-Finkelstein type.

The upshot of all this is simply to justify partially
but precisely the idea
that the geometry of spherical
collapse may be modeled by a moving mirror problem.
In this model the mirror arises at the origin of coordinates
({\it not\/} the horizon); its ``motion'' is an effective representation
of the distortion of space-time in the collapse.  An important feature
left out of the model in its simplest
form is the non-trivial spatial curvature outside the
shell.

\section{Causal Structure of the Mirror Problem}

In the moving mirror problem, we
consider the evolution of a massless scalar field
in 1+1 dimensions subject to the
boundary condition
$$
\phi (z(t), t) ~=~ 0
\eqn\globa
$$
along the mirror trajectory $z = z(t)$.
The scalar field is defined to vanish on the left-hand side of the
mirror.
The effect of the boundary condition is of course that rays incident
on the mirror reflect off it.

As we have discussed,
the mirror plays the role of the origin $r=0$ in
space -- the center of the hole -- in the black hole problem.
Thus reflection off the mirror mimics the propagation of an ingoing
wave to the center and its emergence as an outgoing wave.  The
distortion of space-time -- essentially the
lengthening of space (and shortening of time) near the surface of
the hole --
in during collapse has a dynamical effect similar
to
the
effect of a
{\it rapidly receding\/} mirror in 1+1 dimensional flat space.
Indeed the fundamental effect
is that rays reflected off a rapidly
receding mirror are severely red-shifted --
as are the rays, crucial to Hawking's analysis, which
barely avoid being trapped behind the incipient horizon.

Three types of mirror trajectories
are illustrated in Figures 2-4.

The first trajectory type describes a mirror
that accelerates away
from rest at
$t=0$ and approaches the speed of light asymptotically.
Let the asymptote light-ray be denoted $A$ as in Figure 2.
Since we
are dealing, for simplicity,
with a massless field we may consider only left-moving modes.  Let us
define points 1, 2, 4, 5 as in the Figure, and use the same labels to
distinguish
the rays emanating from these points
at $t=0$.

We see that rays such as 1 and 2, which begin to the left of $A$, intersect
the mirror and propagate out to spatial infinity at the right, denoted
in deference to the black hole interpretation as ${\cal I}_+$.  On the
other hand rays such as 4 and 5, which begin to the right of $A$, never
intersect the mirror.  They propagate to the left infinity, denoted as
${\cal H}_+$.  (This infinity may seem  a little funny from the
point of view of the
metaphorical interpretation of
$z$ as the effective position of the black-hole origin.  The point is
that the effective radial distance from the point of view of wave
propagation is most
appropriately measured in intervals of the tortoise coordinate
$r_*$, which diverges to $-\infty$ at the black hole horizon.
By the way, these rays leave the Figure in finite affine time.)

Now consider the problem of the the evolution of a quantum state
defined for $z>0$ at $t=0$ into the distant future.  Naturally one should
consider first the ground state, defined by the absence of
positive-frequency modes.
It is evident that the time interval between
the arrival of 1 and 2 at a given point in space before they
reflect is much dilated after they reflect.  Thus the frequency of
waves is altered, and negative-frequency wave can acquire positive-frequency
components.  This would be interpreted as the creation of an excited
state on ${\cal I}_+$.  For an appropriate mirror trajectory, as
we shall see,
the state on ${\cal I}_+$ will be a thermal state, with its temperature
related to the rate of acceleration of the mirror.
Clearly all information concerning the state of the field
$\phi$ in region A to the left of $A$ at $t=0$ is propagated to
${\cal I}_+$.

Rays such as 4 and 5 beginning to
the right of $A$ propagate undisturbed to
${\cal H}_+$.
Clearly all information concerning the state of the field
$\phi$ in the region to the right of $A$ at $t=0$ is propagated to
${\cal H}_+$.
If we start with the ground state
on $t=0$, an
observer making measurements on ${\cal H}_+$
also sees his natural
ground state.

Now according to basic principles of quantum mechanics, which of course
are certainly
not contradicted by anything in the simple
model problem under consideration,
a pure state localized to the left of A would propagate into a pure state on
${\cal I}_+$ and a pure state localized to the right of A
would propagate into a pure
state on ${\cal H}_+$.  However, the ground state at $t=0$ is
{\it not\/} pure when restricted either to either side of A.
The positive-frequency
condition forces consideration of modes which
extend over both intervals, and introduces
correlations between these intervals.   Indeed, the two-point function
$\langle \phi (1) \phi (3) \rangle $ at $t=0$, for example,
certainly does not vanish.  Furthermore, this correlation will
propagate in a simple way into the future, introducing correlations
between ${\cal I}_+$ and ${\cal H}_+$.
Thus we should not be shocked to find
a mixed state if we consider ${\cal I}_+$ by itself, without
regard to (tracing over) the state on ${\cal H}_+$.   And this indeed is
what we do find: the correlation functions on ${\cal I}_+$, for the
appropriate trajectory of this type, are {\it precisely\/} thermal,
and therefore
certainly must be described by a mixed state on ${\cal I}_+$.

The phenomenon that may be
a shock to one's intuition
is that
it is correlations between the rich thermal state on
${\cal I}_+$ and the
apparently barren desert on ${\cal H}_+$
which insure purity of the whole.  Thus for example the
expectation value of the energy-momentum tensor vanishes,
and its multi-point correlators are vacuous (\ie\ indistinguishable
from the vacuum),
when restricted to
${\cal H}_+$ -- but its cross-correlators between ${\cal H}_+$ and
${\cal I}_+$ do not vanish.
This peculiar phenomenon, whose existence and nature is made quite
transparent by the foregoing extremely elementary observations,
was noted and emphasized by
Carlitz and Willey [\carlitz].
(However they somewhat obscured the issue by claiming
in effect
that particle creation on ${\cal I}_+$ is uniquely and locally
related
to particle creation on ${\cal H}_+$, which is not
the case.)
It shows in as dramatic fashion as one could desire
that the purity of a big complicated state with gigantic entropy
(in any sense) can be restored at little -- here
actually at {\it zero\/}
-- cost in energy.

Now let us consider the mirror trajectory
depicted in Figure 3, which is the same as the one discussed
for a long interval of time, but such that the mirror eventually
stops accelerating.  Then all rays eventually intersect the
mirror, and get reflected to ${\cal I}_+$.  Thus we obtain on
${\cal I}_+$ a pure state which looks thermal for
an arbitrarily long time.  Of course once the mirror stops accelerating
there is no longer any radiation emitted.  The transition to
zero acceleration can be done smoothly, so that only a small burst
(whose magnitude is essentially independent of the
length of the interval over
which thermal radiation has occurred)
accompanies it.  Thus altogether one finds, similar to the
previous case, that quantum purity comes at a small price.

The situation of Figure 3 is not yet a
model for
complete black hole evaporation.  For although
positive frequencies at late times are indeed reflected into positive
frequencies, and there is no particle production, yet the frequency
is highly red-shifted.  Thus real particles at late times will sense
(in the interpretation of the mirror as the locus of the
origin) a remnant that
delays them for a long time and saps their energy.  It is
left for the reader
to invent witty names for such a remnant.

Finally in Figure 4 we have the situation where the mirror returns
to rest.
Real particles emitted at late times, which intersect
the mirror during its second period of rest, behave as if passing through
the origin of empty space in the analogue problem.
Thus this provides a model for a black hole that evaporates completely.
 From the nature of the construction, any pure initial state evolves into
a pure final state.

\section{Correlation Functions and Renormalized Entropy}

For simple forms of matter, \eg\ a free massless scalar field or
any conformal field theory, the
moving mirror problem is eminently tractable.  Any quantity of interest
may be calculated explicitly.  For example one has for the
correlation function of the fields
$$\eqalign{\Gv& \equiv \bra 0 \phi(1) \phi(2) \ket 0  \cr
        &={1\over 4 \pi} \ln
        {(U_2-U_1+i \delta) (V_2-V_1+i \delta) \over
        (U_2-V_1+i \delta)(V_2-U_1+i \delta)} .\cr }
\eqn\gvacb $$
Indeed this function manifestly satisfies the wave equation with
the correct singularity and the moving mirror boundary condition, and
reduces to the correct vacuum value before the mirror motion begins.
For the energy-momentum tensor describing the emitted radiation one
finds, using \ttrans , \dUdu , \metUV
$$\bra 0 T_{\mu \nu} \ket 0 ={\delta_{\mu u} \delta_{\nu u}
          \over 12 \pi} \sqrt{U'} {d^2 \over du^2} \sqrt{1/ U'}.
         \eqn\tmunu$$

All energy $n$-point functions are
determined by $\bra 0 T_{\mu \nu} \ket 0 $ and $\Gv$.  For example
the energy two-point function
$$C_{\mu\nu ,\alpha\beta}(1,2)\equiv
        G^E_{\alpha \beta, \mu \nu}(1,2)-        G^E_{\alpha \beta}(1)
G^E_{\mu \nu}(2)
        . \eqn\Cdef $$
is evaluated to be
$$\eqalign{ C_{uu,uu}(1,2) &= {1 \over 8 \pi^2}
        {U'(1)^2 U'(2)^2 \over (U(2)-U(1))^4}, \cr C_{vv,vv}(1,2) &= {1
\over 8 \pi^2}
        {1 \over (v(2)-v(1))^4}, \cr C_{uu,vv}(1,2) &= {1 \over 8 \pi^2}
        {U'(1)^2 \over (v(2)-U(1))^4}. }\eqn\Cexpl $$
Not unexpectedly, the correlations diverge for two points connected by a
light-like line in the direction of the energy flux in question.
Note that there
are correlations between leftward and rightward flux, as anticipated.
These correlations are, however, by no means sharply
localized.

One may compare these expressions
to the thermal correlation function (populating
only right-movers) which is easily found to be
$$\eqalign{G_{th}(1,2)=&-{T \over 4} ( |u_1-u_2|+|v_1-v_2| )+ \cr
	&+{1\over 4 \pi} \ln \left[ \left(1-e^{-2 \pi T |u_1-u_2|} \right)
         \left(1-e^{-2 \pi T|v_1-v_2|} \right) \right],}
        \eqn\gthc $$
and to
the two-point
correlation of outward flux:
$$C_{uu,uu}(1,2)={\kappa^4 \over 8 \pi^2}
        {e^{2 \kappa |u_1-u_2|} \over  (e^{\kappa |u_1-u_2|} -1)^4 }~.
        \eqn\twothd $$
There is perfect agreement if we substitute for $U$
the particular trajectory $U \propto -e^{-\kappa u}$, which we shall call the
thermal trajectories. Moreover, with that choice,
$${\partial \over \partial u_1}{\partial \over \partial u_2}
        G_{th}(1,2)=
        {\partial \over \partial u_1}{\partial \over \partial u_2}
        G_{vac}(1,2),
        \eqn\gtheqgvac $$
so that {\it all\/} correlations of outward energy flux will be thermal.
Correlations involving $T_{vv}$ are, of course, not thermal. In fact, we see
from $\Cexpl$ that the two-point correlation
$C_{vv,vv}(1,2)$ for the mirror is the ordinary correlation expected for a
vacuum state.

Now let us discuss the entropy.  Using \sr\ for
an interval bounded by $u_1$ and $u_2$
$$
S_{\rm  ren } ~=~ {1\over 12} \ln {(U_2 - U_1)^2\over
 U_1^\prime U_2^\prime (u_1 -u_2)^2}~.
\eqn\genrenent
$$
Thus for the Schwarzschild trajectory $U=c_1 + c_2 e^{-{1\over 4M}u}$
one finds
$$
S_{\rm ren } ~=~ {1\over 12}
\ln ({(4M)^2\over (u_2 -u_1)^2} (e^{{1\over 8M} (u_2 -u_1)}
- e^{-{1\over 8M} (u_2 -u_1)})^2)~.
\eqn\Schwent
$$
For $u_2 - u_1 \gg 8M$ this is approximately
$$
S_{\rm ren } ~\approx~ {1\over 8\pi M }{\pi \over 6} (u_2- u_1)~.
\eqn\approxent
$$
Remarkably, this purely microscopically defined entropy agrees with
the entropy one would derive by treating the Hawking radiation field
as if it were thermal at temperature $T=1/(8\pi M)$.  This justifies,
at least in the present context, our claim that the renormalized geometric
entropy is a natural concept with a significant physical interpretation.

The renormalized entropy also behaves sensibly for other mirror
trajectories.
Thus for example
it vanishes (trivially) for constant velocity trajectories and
(less trivially) for the trajectories
$$
U~=~ c_1 + {c_2\over u - c_3}
\eqn\rntraj
$$
corresponding to extremal Reissner-Nordstrom black holes, which
emit no Hawking radiation.

However there is a big difference between
the {\it global\/} behavior of the renormalized geometric entropy
and the behavior of the corresponding thermodynamic
entropy of the radiation field, that captures in a quantitative way
the physics discussed in the preceding section.  As we have seen
before (in the discussion around \intent\ ), the
renormalized geometric entropy of an interval
can easily shrink as the interval expands, even
in a quite non-pathological
situation.   Similarly one finds (essentially by the same argument)
that the geometric entropy associated with mirror trajectories which
asymptote to a constant velocity less than c -- including, of course,
0 -- rapidly approaches zero, even though it may have built up in the
thermal fashion discussed above for an arbitrarily long time previously.

On the other hand if a true horizon forms the integrated
renormalized geometric
entropy is generally infinite, since $U_2^\prime \rightarrow 0$ and
 $u_2 \rightarrow \infty$.
An exception occurs if $U_2^\prime \propto u_2^{-2}$ in this limit, which
singles out the mirror trajectories appropriate to extremal black holes.

Evidently it is dangerous to think of microscopic, fine-grained
entropy as a substance which can be measured locally
and once created is never destroyed.  This seems to us to
lessen the force of
one form of the ``information paradox'' for black holes.
Semi-classical calculations of the radiation from black holes
indicate that
their emission is the same as one might expect from an ideal grey body.
It is commonly believed that these calculations are very accurate for
black holes having masses much larger than the Planck mass (and away
from any extremal limit).   This raises a conceptual problem that has
been much discussed, as follows.
One can certainly imagine forming a black hole from
matter in a pure quantum state.  One then finds, in an
approximation which appears accurate,
that it radiates to produce a mixed
state.  Yet the evolution of a pure into a mixed state
would violate the basic
principles of quantum mechanics as they are currently understood.

However,
when
black hole radiance is calculated semiclassically, as the response
of external fields to a given space-time geometry, the calculation is
essentially identical to that for the corresponding mirror problem.
But in the latter problem it is unambiguously clear that the radiance
is only pseudo-thermal, being in a precise sense a
reflection of correlations
present in the vacuum.  (Indeed, this pseudo-thermal character
is already implied by the appearance of the radiance for
free fields, which in principle have no mechanism to enforce
thermal equilibrium!)  It therefore seems most
plausible that the relevant entropy to
consider in assessing the question whether the final
state can be pure, even when going beyond the
semi-classical approximation, is the renormalized geometric entropy.
This is definitely
not an extensive quantity, as presumed in the
preceding, conventional argument which leads to an information paradox.


The important limitation of the moving mirror model is that it
does no justice to the conservation of energy, since the motion of
the mirror is prescribed {\it a priori}\foot{A recent attempt
to incorporate energy conservation in a mirror model can
be found in [\chung]}.  This limitation becomes
particularly serious if we attempt to model complete evaporation
-- {\it i.e}. to bring the mirror to rest.   Indeed
if we define
$$\sqrt{U'}=e^{-g},
	\eqn\defg$$
then we obtain from \tmunu\
the total energy flux radiated after the thermal
period in the form:
$$E={1 \over 12 \pi} \int_{u_e}^{u_r} \left( g'^2 -g'' \right) du.
	\eqn\miraa$$
At $u_e$, the end of the thermal period, we have $g \approx \kappa u_e/2$
and $g' \approx \kappa/2$. If we demand that the  mirror be at rest after
$u=u_r$ (so that $g=g'=0$) and minimize the integral \miraa , the $g''$-term
leads to a constant boundary-term in the variational procedure and a
linearly decreasing $g$ is optimal. The trajectory is therefore of
the thermal form \trath\ (with negative $\kappa$).
The integrated flux
decreases with increasing available time.
If we suppose that deceleration sets in only when the hole
has reached the Planck mass, then the available energy is quite
small and one must stretch out the deceleration process in order
to minimize the radiation.
Carlitz and Willey therefore concluded
that the time interval over
which the mirror gets back to rest, and  space-time returns to normal,
would have be much longer than the lifetime of the black hole.

While this sort of slowly cooling remnant appears to be a logically
consistent possibility, in the absence of a specific mechanism
it seems a sufficiently strange outcome that one is open to
alternatives.  For example, the dynamics at moments when the
formal mirror trajectory is undergoing gargantuan accelerations, which
according to this model yields a gargantuan burst of radiation, is
unlikely to be a valid representation of reality, especially in a theory
with very soft ultraviolet behavior (specifically, string theory).

Although the methods used in this paper clearly are inadequate to
resolve all
the problems connected with black hole quantum mechanics, we do think they
clarify the nature of some of these problems.  In any case,
the
physical significance of renormalized entropy has here been
exemplified concretely in
the analysis of models often used to discuss these problems.
Noteworthy features of this entropy are its lack of additivity and of
local
positivity even in simple non-pathological situations.

{\bf acknowledgement}

We wish to thank C. Callan and H. B. Nielsen for useful discussions.

\endpage

\centerline{\bf figure captions}

{\bf figure 1}
Conformal mapping from a finite strip in a periodic box to
two halflines and further on to the upper and lower boundary of
a strip. The grey areas indicate the development from an early
time that selects the vacuum state at the present.

{\bf figure 2}
A moving mirror configuration with a light--like asymptote:
the mirror accelerates indefinitely asymptoting the speed of light.

{\bf figure 3}
A moving mirror with a time-like asymptote:
The mirror stops accelerating after a finite time
and moves at constant velocity.

{\bf figure 4}
A moving mirror that eventually comes to rest.

\endpage

\refout

\end